\documentclass[12pt]{article}
\pdfoutput=1
\usepackage[colorlinks,linkcolor=Blue,citecolor=Blue,bookmarks,bookmarksnumbered]{hyperref}
\usepackage[scaled=0.85]{helvet}
\usepackage{amsmath,amssymb,accents,mathrsfs,XoohmE}
\usepackage{graphicx,color}
\graphicspath{{Figures/}}
\usepackage{booktabs}
\usepackage{multirow}
\usepackage{placeins}
\usepackage{amsmath}
\usepackage{subfigure}

\usepackage{XoohmE}

\definecolor{Green}  {rgb}{0.10,0.70,0.10} 
\definecolor{Orange} {rgb}{1.00,0.50,0.15} 
\definecolor{Red}    {rgb}{0.90,0.00,0.12} 
\definecolor{Purple} {rgb}{0.50,0.25,0.55} 
\definecolor{Turque} {rgb}{0.00,0.65,0.85} 
\definecolor{Blue}   {rgb}{0.00,0.00,1.00} 
\definecolor{Magenta}{rgb}{1.00,0.00,1.00} 
\definecolor{Gold}   {rgb}{1.00,0.75,0.25} 
\definecolor{Seaweed}{rgb}{0.01,0.24,0.09} 
\definecolor{Brown}  {rgb}{0.43,0.26,0.32} 
\definecolor{grey1}  {rgb}{0.20,0.20,0.20} 
\definecolor{grey2}  {rgb}{0.40,0.40,0.40} 
\definecolor{grey3}  {rgb}{0.60,0.60,0.60} 
\definecolor{grey4}  {rgb}{0.80,0.80,0.80} 
\definecolor{grey5}  {rgb}{0.90,0.90,0.90} 
\def\C#1#2{{\ifcase#1\or
             \color{Green}\or \color{Orange}\or \color{Red}\or
              \color{Purple}\or \color{Turque}\or \color{Blue}\or
               \color{Magenta}\or \color{Gold}\or \color{Seaweed}\or
                \color{Brown}\or\color{grey1}\or\color{grey2}\or
                 \color{grey3}\else\color{grey4}\fi#2}}

\definecolor{Slate} {rgb}{0.00,0.45,0.55}

\usepackage{color} 
\usepackage{listings} 
\usepackage{setspace} 
 
\definecolor{Code}{rgb}{0,0,0} 
\definecolor{Decorators}{rgb}{0.5,0.5,0.5} 
\definecolor{Numbers}{rgb}{0.5,0,0} 
\definecolor{MatchingBrackets}{rgb}{0.25,0.5,0.5} 
\definecolor{Keywords}{rgb}{0,0,1} 
\definecolor{self}{rgb}{0,0,0} 
\definecolor{Strings}{rgb}{0,0.63,0} 
\definecolor{Comments}{rgb}{0,0.63,1} 
\definecolor{Backquotes}{rgb}{0,0,0} 
\definecolor{Classname}{rgb}{0,0,0} 
\definecolor{FunctionName}{rgb}{0,0,0} 
\definecolor{Operators}{rgb}{0,0,0} 
\definecolor{Background}{rgb}{0.98,0.98,0.98}

\lstdefinelanguage{Python}{ 
numbers=left, 
numberstyle=\footnotesize, 
numbersep=1em, 
xleftmargin=1em, 
framextopmargin=2em, 
framexbottommargin=2em, 
showspaces=false, 
showtabs=false, 
showstringspaces=false, 
frame=l, 
tabsize=4, 
basicstyle=\ttfamily\small\setstretch{1}, 
backgroundcolor=\color{Background}, 
commentstyle=\color{Comments}\slshape, 
stringstyle=\color{Strings}, 
morecomment=[s][\color{Strings}]{"""}{"""}, 
morecomment=[s][\color{Strings}]{'''}{'''}, 
morekeywords={import,from,class,def,for,while,if,is,in,elif,else,not,and,or,print,break,continue,return,True,False,None,access,as,,del,except,exec,finally,global,import,lambda,pass,print,raise,try,assert}, 
keywordstyle={\color{Keywords}\bfseries}, 
morekeywords={[2]@invariant,pylab,numpy,np,scipy}, 
keywordstyle={[2]\color{Decorators}\slshape}, 
emph={self}, 
emphstyle={\color{self}\slshape}, 
}

\def\rI{{\rm I}}
\def\rJ{{\rm J}}

\def\rL{{\rm L}}
\def\rR{{\rm R}}

\def\fracm#1#2{\hbox{\large{${\frac{{#1}}{{#2}}}$}}}

\def\be{\begin{equation}}
\def\ee{\end{equation}}
\newcommand{\bea}{\begin{eqnarray}}
\newcommand{\eea}{\end{eqnarray}}
\newcommand{\ena}{\end{eqnarray}}

\def\pp{{\mathchoice
          {
              \kern 1pt%
              \raise 1pt
              \vbox{\hrule width5pt height0.4pt depth0pt
                    \kern -2pt
                    \hbox{\kern 2.3pt
                          \vrule width0.4pt height6pt depth0pt
                          }
                    \kern -2pt
                    \hrule width5pt height0.4pt depth0pt}%
                    \kern 1pt
           }
            {
              \kern 1pt%
              \raise 1pt
              \vbox{\hrule width4.3pt height0.4pt depth0pt
                    \kern -1.8pt
                    \hbox{\kern 1.95pt
                          \vrule width0.4pt height5.4pt depth0pt
                          }
                    \kern -1.8pt
                    \hrule width4.3pt height0.4pt depth0pt}%
                    \kern 1pt
            }
            {
              \kern 0.5pt%
              \raise 1pt
              \vbox{\hrule width4.0pt height0.3pt depth0pt
                    \kern -1.9pt  
                    \hbox{\kern 1.85pt
                          \vrule width0.3pt height5.7pt depth0pt
                          }
                    \kern -1.9pt
                    \hrule width4.0pt height0.3pt depth0pt}%
                    \kern 0.5pt
            }
            {
              \kern 0.5pt%
              \raise 1pt
              \vbox{\hrule width3.6pt height0.3pt depth0pt
                    \kern -1.5pt
                    \hbox{\kern 1.65pt
                          \vrule width0.3pt height4.5pt depth0pt
                          }
                    \kern -1.5pt
                    \hrule width3.6pt height0.3pt depth0pt}%
                    \kern 0.5pt
            }
        }}

\def\mm{{\mathchoice
                       {
                             \kern 1pt
               \raise 1pt    \vbox{\hrule width5pt height0.4pt depth0pt
                                  \kern 2pt
                                  \hrule width5pt height0.4pt depth0pt}
                             \kern 1pt}
                       {
                            \kern 1pt
               \raise 1pt \vbox{\hrule width4.3pt height0.4pt depth0pt
                                  \kern 1.8pt
                                  \hrule width4.3pt height0.4pt depth0pt}
                             \kern 1pt}
                       {
                            \kern 0.5pt
               \raise 1pt
                            \vbox{\hrule width4.0pt height0.3pt depth0pt
                                  \kern 1.9pt
                                  \hrule width4.0pt height0.3pt depth0pt}
                            \kern 1pt}
                       {
                           \kern 0.5pt
             \raise 1pt  \vbox{\hrule width3.6pt height0.3pt depth0pt
                                  \kern 1.5pt
                                  \hrule width3.6pt height0.3pt depth0pt}
                           \kern 0.5pt}
                       }}

\def\ad{{\kern0.5pt
                   \alpha \kern-5.05pt \raise5.8pt\hbox{$\textstyle.$}\kern
0.5pt}}

\def\bd{{\kern0.5pt
                   \beta \kern-5.05pt \raise5.8pt\hbox{$\textstyle.$}\kern
0.5pt}}

\def\qd{{\kern0.5pt
                   q \kern-5.05pt \raise5.8pt\hbox{$\textstyle.$}\kern
0.5pt}}
\def\Dot#1{{\kern0.5pt
     {#1} \kern-5.05pt \raise5.8pt\hbox{$\textstyle.$}\kern
0.5pt}}

\catcode`@=11
\def\un#1{\relax\ifmmode\@@underline#1\else
        $\@@underline{\hbox{#1}}$\relax\fi}
\catcode`@=12

\def\a{\alpha}
\def\b{\beta}
\def\c{\chi}
\def\d{\delta}

\def\g{\gamma}

\def\i{\iota}
\def\j{\psi}
\def\k{\kappa}
\def\l{\lambda}

\def\o{\omega}

\def\r{\rho}
\def\s{\sigma}
\def\t{\tau}

\def\L{\Lambda}
\def\O{\Omega}
\def\P{\Pi}

\def\S{\Sigma}

\def\ca{{\cal A}}

\def\cf{{\cal F}}

\def\dslash{\not{\hbox{\kern-2pt $\partial$}}}
\def\Dslash{\not{\hbox{\kern-4pt $D$}}}
\def\pslash{\not{\hbox{\kern-2.3pt $p$}}}
 \newtoks\slashfraction
 \slashfraction={.13}
 \def\slash#1{\setbox0\hbox{$ #1 $}
 \setbox0\hbox to \the\slashfraction\wd0{\hss \box0}/\box0 }
 
\def\kcr{{\hbox{\ro \char'170}}}                
\def\ktl{{\hbox{\ro \char'170}}}        
\def\ktr{{\hbox{\ro \char'170}}}        
\def\kbl{{\hbox{\ro \char'170}}}        
\def\kbr{{\hbox{\ro \char'170}}}        

\def\plpl{\raise-2pt\hbox{$\raise3pt\hbox{$_+$}\hskip-6.67pt\raise0.0pt
\hbox{$^+$}\hskip 0.01pt$}}
\def\mimi{\raise-2pt\hbox{$\raise3pt\hbox{$_-$}\hskip-6.67pt\raise0.0pt
\hbox{$^-$}\hskip 0.01pt$}} 

\def\bo{{\raise.15ex\hbox{\large$\Box$}}}               
\def\TH{{\raise.2ex\hbox{$\displaystyle \bigodot$}\mskip-4.7mu \llap H \;}}
\def\face{{\raise.2ex\hbox{$\displaystyle \bigodot$}\mskip-2.2mu \llap {$\ddot
        \smile$}}}                                      

\def\dt#1{\on{\hbox{\bf .}}{#1}}                
\def\Dot#1{\dt{#1}}

\def\Hat#1{\widehat{#1}}                        
\def\leftrightarrowfill{$\mathsurround=0pt \mathord\leftarrow \mkern-6mu
        \cleaders\hbox{$\mkern-2mu \mathord- \mkern-2mu$}\hfill
        \mkern-6mu \mathord\rightarrow$}
\def\dvec#1{\vbox{\ialign{##\crcr
        \leftrightarrowfill\crcr\noalign{\kern-1pt\nointerlineskip}
        $\hfil\displaystyle{#1}\hfil$\crcr}}}           
\def\dt#1{{\buildrel {\hbox{\LARGE .}} \over {#1}}}     

\def\fracm#1#2{\hbox{\large{${\frac{{#1}}{{#2}}}$}}}
\def\sfrac#1#2{{\vphantom1\smash{\lower.5ex\hbox{\small$#1$}}\over
        \vphantom1\smash{\raise.4ex\hbox{\small$#2$}}}} 
\def\bfrac#1#2{{\vphantom1\smash{\lower.5ex\hbox{$#1$}}\over
        \vphantom1\smash{\raise.3ex\hbox{$#2$}}}}       
\def\afrac#1#2{{\vphantom1\smash{\lower.5ex\hbox{$#1$}}\over#2}}    

\let\bm\relax
\newcommand{\bm}[1]{{\boldsymbol{#1}}}

\def\ad{{\dot{\alpha}}}
\def\bd{{\dot{\beta}}}

 \font\rOpe=cmsy10                        
 \def\ktl{{\hbox{\rOpe\char'170}}}        
 \def\kbl{{\hbox{\rOpe\char'170}}}        
 \def\kcr{{\reflectbox{\rOpe\char'170}}}        
 \def\ktr{{\reflectbox{\rOpe\char'170}}}        
 \def\kbr{{\reflectbox{\rOpe\char'170}}}        
 \def\Border{\vbox{\hsize0pt
        \setlength{\unitlength}{1mm}
        \newcount\xco
        \newcount\yco
        \xco=-21
        \yco=12
        \begin{picture}(0,0)(-7.5,0)
        \put(\xco,\yco){$\ktl$}
        \advance\yco by-1
        {\loop
        \put(\xco,\yco){$\kcr$}
        \advance\yco by-2
        \ifnum\yco>-240
        \repeat
        \put(\xco,\yco){$\kbl$}}
        \xco=170
        \yco=12
        \put(\xco,\yco){$\ktr$}
        \advance\yco by-1
        {\loop
        \put(\xco,\yco){$\kcr$}
        \advance\yco by-2
        \ifnum\yco>-240
        \repeat
        \put(\xco,\yco){$\kbr$}}
        \put(-19.5,13){\scalebox{.6065}{%
         University of Maryland Center for String and Particle  Theory \&\ Physics Department%
        |University of Maryland Center for String and Particle  Theory \&\ Physics Department}}
        \put(-19.5,-241.5){\scalebox{.5835}{%
         ****University of Maryland * Center for String and
         Particle  Theory* Physics Department****University of Maryland *Center
        for String and Particle  Theory* Physics Department}}
        \end{picture}
        \par\vskip-8mm}}
\definecolor{UMred}{rgb}{.9,.05,.2}
\definecolor{HUblue}{rgb}{.0,.3,.7}

\definecolor{Red}    {rgb}{0.90,0.00,0.12} 
\definecolor{Blue}   {rgb}{0.00,0.00,1.00} 
\definecolor{Green}  {rgb}{0.10,0.70,0.10} 
\definecolor{Turque} {rgb}{0.00,0.65,0.85} 
\definecolor{Orange} {rgb}{1.00,0.50,0.15} 
\definecolor{Magenta}{rgb}{1.00,0.00,1.00} 
\definecolor{Gold}   {rgb}{1.00,0.75,0.25} 
\definecolor{Seaweed}{rgb}{0.01,0.24,0.09} 
\definecolor{Purple} {rgb}{0.50,0.25,0.55} 
\definecolor{Brown}  {rgb}{0.43,0.26,0.32} 
\definecolor{grey1}  {rgb}{0.20,0.20,0.20} 
\definecolor{grey2}  {rgb}{0.40,0.40,0.40} 
\definecolor{grey3}  {rgb}{0.60,0.60,0.60} 
\definecolor{grey4}  {rgb}{0.80,0.80,0.80} 
\definecolor{grey5}  {rgb}{0.90,0.90,0.90} 
\def\C#1#2{{\ifcase#1\or
             \color{Red}\or \color{Green}\or \color{Blue}\or\
              \color{Turque}\or \color{Orange}\or \color{Magenta}\or 
               \color{Gold}\or \color{Seaweed}\or \color{Purple}\or
                \color{Brown}\or\color{grey1}\or\color{grey2}\or
                 \color{grey3}\else\color{grey4}\fi#2}}

\definecolor{Slate} {rgb}{0.00,0.45,0.55}

\newdimen\parshift\parshift=\parindent
\catcode`@=11
 \long\def\@footnotetext#1{\insert\footins{\reset@font\footnotesize
           \interlinepenalty\interfootnotelinepenalty\splittopskip%
            \footnotesep\splitmaxdepth\dp\strutbox\floatingpenalty\@MM%
             \hsize\columnwidth\addtolength{\hsize}{-2\parindent}
              \@parboxrestore\protected@edef\@currentlabel%
              {\csname p@footnote\endcsname\@thefnmark}%
                \color@begingroup%
                 \@makefntext{\rule\z@\footnotesep\ignorespaces#1%
                  \@finalstrut\strutbox}%
                \color@endgroup}}
 \long\def\@makefntext#1{\hglue\parshift%
           \vbox{\noindent\baselineskip=11pt plus.5pt minus.5pt\hb@xt@0em{\hss\@makefnmark\kern1pt}#1}}
\catcode`@=12

\newskip\humongous \humongous=0pt plus 1000pt minus 1000pt
\def\caja{\mathsurround=0pt}
\def\eqalign#1{\,\vcenter{\openup2\jot \caja
        \ialign{\strut \hfil$\displaystyle{##}$&$
        \displaystyle{{}##}$\hfil\crcr#1\crcr}}\,}
\newif\ifdtup

\makeatletter
\def\section{\@startsection{section}{1}{\z@}
        {3ex plus-1ex minus-.2ex}{1pt plus1pt}{\large\sf\bfseries\boldmath}}
\def\subsection{\@startsection{subsection}{2}{\z@}
         {1.5ex plus-1ex minus-.2ex}{0.01pt plus1pt}{\sf\slshape}}
\def\subsubsection{\@startsection{subsubsection}{3}{\z@}
          {1.5ex plus-1ex minus-.2ex}{0.01pt plus0.2pt}{\sf\boldmath}}
\def\paragraph{\@startsection{paragraph}{4}{\z@}
           {.75ex \@plus.5ex \@minus.2ex}{-2mm}{\sf\bfseries\boldmath}}
\makeatother

 \allowdisplaybreaks
 \seceq

\usepackage{lipsum}
\usepackage{listings}
\definecolor{MyDarkGreen}{rgb}{0.0,0.4,0.0} 
\usepackage[enableskew,vcentermath]{youngtab}

\definecolor{Hey}{rgb}{.9,.05,.4}
\definecolor{orange}{rgb}{1,.5,0}
\definecolor{plum}{rgb}{.4,0,.6}
\definecolor{R}{rgb}{1,0,0}
\definecolor{G}{rgb}{0.1,0.7,0}
\definecolor{B}{rgb}{0,0,1}
\begin{document}

\thispagestyle{empty}
\noindent{\small
\hfill{  \\ 
$~~~~~~~~~~~~~~~~~~~~~~~~~~~~~~~~~~~~~~~~~~~~~~~~~~~~~~~~~~~~~~~~~$
$~~~~~~~~~~~~~~~~~~~~~~~~~~~~~~~~~~~~~~~~~~~~~~~~~~~~~~~~~~~~~~~~~$
{}
}}
\vspace*{0mm}
\begin{center}
{\large \bf
 \hskip1.5in
A Note On \newline
Exemplary Off-Shell Constructions Of 4D, 
$\bm {\cal N}$ = 2 Supersymmetry Representations
 \\[2pt]
}   \vskip0.3in
{\large {
$~~~~~~~~$
Devin D.\ Bristow\footnote{devin.bristow@pepperdine.edu}$^{,a}$,
John H.\ Caporaletti\footnote{jcapor@terpmail.umd.edu}$^{,b}$,
Aleksander J.\ Cianciara\footnote{aleksander${}_-$cianciara@brown.edu}$^{,c,d}$,
\newline
$~~~~~~~~~~~~~$
S.\ James Gates, Jr.\footnote{sylvester${}_-$gates@brown.edu}$^{,c,d}$, 
Delina Levine\footnote{dmlevine@terpmail.umd.edu}$^{,b}$, and
Gabriel Yerger\footnote{Gabrielyerger@gmail.com}$^{,b}$ $~~~~~~$
}}
\\*[8mm]
\emph{
\centering
$^{a}$Pepperdine University, Natural Science Division
\\[1pt]
Malibu, CA 90263, USA,
\\[12pt]
$^{b} $Department of Physics, University of Maryland,
\\[1pt]
College Park, MD 20742-4111, USA
\\[12pt]
$^{c}$Brown University, Department of Physics,
\\[1pt]
Box 1843, 182 Hope Street, Barus \& Holley 545,
Providence, RI 02912, USA,
\\[4pt] and \\[4pt]
$^{d}$Brown Center for Theoretical Physics, 
340 Brook Street, Barus Hall,
Providence, RI 02912, USA,
}
 \\*[15mm]
{ ABSTRACT}\\[4mm]
\parbox{142mm}{\parindent=2pc\indent\baselineskip=14pt plus1pt
We continue the search for rules that govern when off-shell 4D,
$\cal N$ = 1 supermultiplets can be combined to form off-shell 4D,
$\cal N$ = 2 supermultiplets. We study the ${\mathbb S}_8$ permutations
and Height Yielding Matrix Numbers (HYMN) embedded within the adinkras 
that correspond to these putative off-shell 4D, $\cal N$ = 2 supermultiplets. 
Even though the HYMN definition was designed to distinguish between the 
raising and lowering of nodes in one dimensional valise supermultiplets, 
they are shown to accurately select out which combinations of off-shell 4D, 
$\cal N$ = 1 supermultiplets correspond to off-shell 4D, $\cal N$ = 2 
supermultiplets.  Only the combinations of the chiral + vector and chiral + 
tensor are found to have valises in the same class.  This is consistent with 
the well known structure of  4D, $\cal N$ = 2 supermultiplets.}
 \end{center}
\vfill
\vfill
\clearpage
%

%
\section{Introduction}
\label{sec:NTR0}

There remain a few simple, unanswered questions in supersymmetry (SUSY).  The simplest form of one such class of questions is,  

``Given minimal representations of off-shell 4D, $\cal N $ = 1 supermultiplets, what combinations of these can be used as a basis for forming off-shell 4D, $\cal N $ = 2 supermultiplets?"

We inaugurated our studies on this in a 2014 \cite{adnkKyeoh} research investigation. This current work represents a continuation along that line. In particular, we propose to use the information contained in the adinkra \cite{Adnk1} projections of these supermultiplets to one-dimensional supersymmetrical systems in order to answer this question. 

Before turning to the 4D, ${\cal N} $ = 2 theories it is useful to recall our 4D, ${\cal N}$ = 1 supermultiplets.  Every off-shell 4D, ${\cal N} $ = 1 supermultiplet
reduced to one dimension leads to a set of matrices that satisfy the Garden algebra \cite{Garden}
\begin{align}
\begin{split}
{\bm {\rL}}{}_{{}_{\rI}} \, {\bm {\rR}}{}_{{}_{\rJ}} ~+~ {\bm {\rL}}{}_{{}_{\rJ}} \, {\bm {\rR}}{}_{{}_{\rI}}
& ~=~ 2\,  \d{}_{{}_{\rm {I \, J}}} \, {\bm {\rm{I}}}{}_{\rm d}  ~~~, ~~~
{\bm {\rR}}{}_{{}_{\rI}} \, {\bm {\rL}}{}_{{}_{\rJ}} ~+~ {\bm {\rR}}{}_{{}_{\rJ}} \, {\bm {\rL}}{}_{{}_{\rI}}
~=~ 2\,  \d{}_{{}_{\rm {I \, J}}} \, {\bm {\rm{I}}}{}_{\rm d} ~~~.
\end{split}
\label{eq:GAlg1}
\end{align}
where the index I takes on values 1, $\dots $, 4 and d = 4p, where p is a non-negative integer
and equal to one for minimal representations.  Each of the matrices ${\bm {\rL}}_{\, {\rI}}$ takes 
the form
\begin{equation}
 {\bm {\rL}}_{\,  {\rI}} 
 ~=~  {\bm {\cal S}}_{\,  {\rI}}  \, {\bm {\cal P}}_{\,  {\rI}}  ~~~,
\label{eq:aas1}
\end{equation}
for each {\em {fixed}} value of I. Further, each matrix $ {\bm {\cal S}}_{\,  {\rI}}$ is diagonal and
squares to the identity, and each matrix ${\bm {\cal P}}_{\,  {\rI}}$ describes a permutation.
Thus, reduction to 1D provides a prescription for mapping supermultiplets onto elements of
the permutation group.

%

\section{Potentially `Colorful' Off-shell 4D, $\bm {\cal N} $ = 2 SUSY Multiplets}
\label{sec:ColoR}

In the work of \cite{adnkKyeoh}, by starting from pairs of the minimal off-shell
4D, $\bm {\cal N} $ = 1 chiral, tensor, and vector supermultiplets and their free
actions, a second potential supersymmetry operator was constructed for the six different
choices of pairings of the supermultiplets.  A ``representation label'' ${({\cal R})}$ was
introduced to describe the six pairings: $(CC)$, $(CT)$, $(CV)$,  $(TT)$, $(TV)$, 
and $(VV)$, where for example $(CC)$ would refer to the Chiral + Chiral supermultiplet and $(TV)$ would refer to the Tensor + Vector supermultiplet. For each value of the representation label the pairs of supermultiplets were reduced to one dimensional theories with extended supersymmetry. This
led to eight ${\bm {\rm L}}$ matrices for each pairing.  In the following, we list
the ${\bm {\rm L}}$ matrices in forms that can readily be used to generate the
factorization shown in Eq.\ (\ref{eq:aas1}). 

We factor \cite{permutadnk} the signed permutation matrices $$\left(\mathrm{L}_{\mathrm{I}}\right)_{i}^{\hat{k}}=\left(\mathcal{S}^{(\mathrm{I})}\right)_{i}^{~\hat{\ell}}\left(\mathcal{P}_{(\mathrm{I})}\right)_{\hat{\ell}}^{~\hat{k}}$$ for each  fixed $I = 1,2,...,N$ where the first factor corresponds to a $d \times d$ diagonal matrix with only $\pm 1$ entries, and the second corresponds to a matrix representation of the permutation of d objects. The signed factor can then be rewritten in a (reversed) binary notation where 

$$\left(\mathcal{S}^{(\mathrm{I})}\right)_{i}^{\hat{\ell}}=\left[\begin{array}{cccc}
(-1)^{b_{1}} & 0 & 0 & \cdots \\
0 & (-1)^{b_{2}} & 0 & \cdots \\
0 & 0 & (-1)^{b_{2}} & \cdots \\
\vdots & \vdots & \vdots & \ddots
\end{array}\right] \quad \leftrightarrow \quad\left(\mathcal{R}_{\mathrm{I}}=\sum_{i=1}^{\mathrm{d}} b_{i} 2^{i-1}\right)_{b}$$

Using this notation, ${({\cal R})}$ = $(CC)$, the ${\bm {\rm L}}$-matrices can be presented as:
\be
\eqalign{
{\bm {\rm L}}{}_{1}  ~=~ \left[\begin{array}{cc}
(10)_b   (243) &  0  \\
0 &   (10)_b    (243)  \\
\end{array}\right]   ~~&,~~
{\bm {\rm L}}{}_{2}  ~=~ \left[\begin{array}{cc}
(12)_b   (123) &  0  \\
0 &   (12)_b    (123)  \\
\end{array}\right]  ~~,
\cr
{\bm {\rm L}}{}_{3}  ~=~ \left[\begin{array}{cc}
(6)_b   (134) &  0  \\
0 &   (6)_b    (134)  \\
\end{array}\right]   ~~~~~&,~~
{\bm {\rm L}}{}_{4}  ~=~ \left[\begin{array}{cc}
(0)_b   (142) &  0  \\
0 &   (0)_b    (142)  \\
\end{array}\right]  ~~~~~,  \cr
 {\bm {\rm L}}{}_{5}  ~=~ \left[\begin{array}{cc}
0 &  (15)_b   (243)  \\
(0)_b   (243) &   0  \\
\end{array}\right]   ~~&,~~
{\bm {\rm L}}{}_{6}  ~=~ \left[\begin{array}{cc}
0  &  (9)_b   (123)  \\
(6)_b    (123) &   0  \\
\end{array}\right]  ~~~~~, \cr
{\bm {\rm L}}{}_{7}  ~=~ \left[\begin{array}{cc}
0 &  (3)_b   (134)  \\
(12)_b   (134) &   0  \\
\end{array}\right]   ~~&,~~
{\bm {\rm L}}{}_{8}  ~=~ \left[\begin{array}{cc}
0  &  (5)_b   (142)  \\
(10)_b    (142) &   0  \\
\end{array}\right]  ~~~. }  \label{eq:CC}
\ee

For ${({\cal R})}$ = $(CT)$, the ${\bm {\rm L}}$-matrices can be presented as:
\be \eqalign{
{\bm {\rm L}}{}_{1}  ~=~ \left[\begin{array}{cc}
(10)_b   (243) &  0  \\
0 &   (14)_b    (234)  \\
\end{array}\right]   ~~&,~~
{\bm {\rm L}}{}_{2}  ~=~ \left[\begin{array}{cc}
(12)_b   (123) &  0  \\
0 &   (4)_b    (124)  \\
\end{array}\right] ~~~~~, \cr
{\bm {\rm L}}{}_{3}  ~=~ \left[\begin{array}{cc}
(6)_b   (134) &  0  \\
0 &   (8)_b    (132)  \\
\end{array}\right]   ~~~~~&,~~
{\bm {\rm L}}{}_{4}  ~=~ \left[\begin{array}{cc}
(0)_b   (142) &  0  \\
0 &   (2)_b    (143)  \\
\end{array}\right]  ~~~~\,~~,  \cr
 {\bm {\rm L}}{}_{5}  ~=~ \left[\begin{array}{cc}
0 &  (11)_b   (243)  \\
(0)_b   (234) &   0  \\
\end{array}\right]   ~~\,~&,~~
{\bm {\rm L}}{}_{6}  ~=~ \left[\begin{array}{cc}
0  &  (13)_b   (123)  \\
(10)_b    (124) &   0  \\
\end{array}\right]  ~~~~,   \cr
{\bm {\rm L}}{}_{7}  ~=~ \left[\begin{array}{cc}
0 &  (7)_b   (134)  \\
(6)_b   (132) &   0  \\
\end{array}\right]   ~~~~\,~&,~~
{\bm {\rm L}}{}_{8}  ~=~ \left[\begin{array}{cc}
0  &  (1)_b   (142)  \\
(12)_b    (143) &   0  \\
\end{array}\right]  ~~~~~.
} \label{eq:CT}
\ee

For ${({\cal R})}$ = $(CV)$, the ${\bm {\rm L}}$-matrices can be presented as:
\be
\eqalign{
{\bm {\rm L}}{}_{1}  ~=~ \left[\begin{array}{cc}
(10)_b   (243) &  0  \\
0 &   (10)_b    (1243)  \\
\end{array}\right]   ~~&,~~
{\bm {\rm L}}{}_{2}  ~=~ \left[\begin{array}{cc}
(12)_b   (123) &  0  \\
0 &   (12)_b    (23)  \\
\end{array}\right] ~~~~~, \cr
{\bm {\rm L}}{}_{3}  ~=~ \left[\begin{array}{cc}
(6)_b   (134) &  0  \\
0 &   (0)_b    (14)  \\
\end{array}\right]   ~~~~~~~~&,~~
{\bm {\rm L}}{}_{4}  ~=~ \left[\begin{array}{cc}
(0)_b   (142) &  0  \\
0 &   (6)_b    (1342)  \\
\end{array}\right]  ~~~~~,   \cr
 {\bm {\rm L}}{}_{5}  ~=~ \left[\begin{array}{cc}
0 &  (2)_b   (243)  \\
(13)_b   (1243) &   0  \\
\end{array}\right]   ~~~~&,~~
{\bm {\rm L}}{}_{6}  ~=~ \left[\begin{array}{cc}
0  &  (4)_b   (123)  \\
(11)_b    (23) &   0  \\
\end{array}\right]  ~~~~~~,  \cr
{\bm {\rm L}}{}_{7}  ~=~ \left[\begin{array}{cc}
0 &  (14)_b   (134)  \\
(7)_b   (14) &   0  \\
\end{array}\right]   ~~~~~~\,~&,~~
{\bm {\rm L}}{}_{8}  ~=~ \left[\begin{array}{cc}
0  &  (8)_b   (142)  \\
(1)_b    (1342) &   0  \\
\end{array}\right]  ~~\,~~.
}  \label{eq:CV}
\ee

For ${({\cal R})}$ = $(TT)$, the ${\bm {\rm L}}$-matrices can be presented as:
\be
\eqalign{
{~~~~~~}
{\bm {\rm L}}{}_{1}  ~&=~ \left[\begin{array}{cc}
n_+ (14)_b    (234) &  0  \\
0 &  m_+ (14)_b    (234)  \\
\end{array}\right]   ~~,~~  
{\bm {\rm L}}{}_{2}  ~
=~ \left[\begin{array}{cc}
n_+ (4)_b    (124) &  0  \\
0 & m_+  (4)_b    (124)  \\
\end{array}\right] ~~, \cr
{\bm {\rm L}}{}_{3}  ~&=~ \left[\begin{array}{cc}
n_+ (8)_b    (132) &  0  \\
0 &  m_+ (8)_b    (132)  \\
\end{array}\right]   ~~~~~,~~  
{\bm {\rm L}}{}_{4}  ~
=~ \left[\begin{array}{cc}
n_+ (2)_b    (143) &  0  \\
0 &  m_+ (2)_b    (143)  \\
\end{array}\right]  ~~,   \cr
 {\bm {\rm L}}{}_{5}  ~&=~ \left[\begin{array}{cc}
 0 &  n_- (14)_b   (234)  \\
m_- (14)_b   (234) &   0  \\
\end{array}\right]   ~~,~~ 
{\bm {\rm L}}{}_{6}  ~
=~ \left[\begin{array}{cc}
0  & n_- (4)_b    (124)  \\
m_- (4)_b    (124) &   0  \\
\end{array}\right]  ~~,  \cr 
{\bm {\rm L}}{}_{7}  ~&=~ \left[\begin{array}{cc}
0 & n_- (8)_b   (132)  \\
m_- (8)_b   (132) &   0  \\
\end{array}\right]   ~~~~~,~~ 
{\bm {\rm L}}{}_{8}  ~
=~ \left[\begin{array}{cc}
0  & n_- (2)_b    (143)  \\
m_- (2)_b    (143) &   0  \\
\end{array}\right]  ~~,
}  \label{eq:TT}
\ee
where
\be \eqalign{
m_{\pm} ~&=~ {\sqrt 2} \, \cos \left[ \fracm {\pi}4 (2 m \mp 1) \right] ~~~,
~~~
n_{\pm} ~=~ {\sqrt 2} \, \cos \left[ \fracm {\pi}4 (2 n \mp 1) \right]
~~~.
 }   \label{eq:Fax}
\ee

For ${({\cal R})}$ = $(TV)$, the ${\bm {\rm L}}$-matrices can be presented as:
\be
\eqalign{
{~~~~~}
{\bm {\rm L}}{}_{1}  ~&=~ \left[\begin{array}{cc}
n_+ (14)_b    (234)&  0  \\
0 & m_+ (10)_b    (1243)    \\
\end{array}\right]   ~~,~~  
{\bm {\rm L}}{}_{2}  ~
=~ \left[\begin{array}{cc}
 n_+   (4)_b    (124) &  0  \\
0 & m_+ (12)_b    (23)   \\
\end{array}\right]  \,~~, \cr
{\bm {\rm L}}{}_{3}  ~&=~ \left[\begin{array}{cc}
n_+ (8)_b    (132)  &  0  \\
0 & m_+ (0)_b    (14)  \\
\end{array}\right]   ~~~~~~~~,~~  
{\bm {\rm L}}{}_{4}  ~ 
=~ \left[\begin{array}{cc}
 n_+ (2)_b    (143) &  0  \\
0 & m_+ (6)_b    (1342)  \\
\end{array}\right]  ~,   \cr
 {\bm {\rm L}}{}_{5}  ~&=~ \left[\begin{array}{cc}
 0 &n_- (14)_b    (234)  \\
 m_- (10)_b   (1243)  &   0  \\
\end{array}\right]   ~~,~~ 
{\bm {\rm L}}{}_{6}  ~
=~ \left[\begin{array}{cc}
0  & n_-  (4)_b    (124) \\
m_- (12)_b    (23)  &   0  \\
\end{array}\right]  ~\,~,  \cr
{\bm {\rm L}}{}_{7}  ~&=~ \left[\begin{array}{cc}
0 & n_- (8)_b    (132)  \\
m_- (0)_b   (14) &   0  \\
\end{array}\right]   ~~~~~~~~,~~  
{\bm {\rm L}}{}_{8}  ~
=~ \left[\begin{array}{cc}
0  & n_-  (2)_b    (143) \\
 m_- (6)_b    (1342) &   0  \\
\end{array}\right]  ~.
}   \label{eq:TV}
\ee

For ${({\cal R})}$ = $(VV)$, the ${\bm {\rm L}}$-matrices can be presented as:
\be
\eqalign{
{\bm {\rm L}}{}_{1}  ~&=~ \left[\begin{array}{cc}
n_+ (10)_b    (1243) &  0  \\
0 &  m_+ (10)_b    (1243)  \\
\end{array}\right]   ~~,~~  
{\bm {\rm L}}{}_{2}  ~
=~ \left[\begin{array}{cc}
n_+ (12)_b    (23) &  0  \\
0 & m_+  (12)_b    (23)  \\
\end{array}\right] \,~~, \cr
{\bm {\rm L}}{}_{3}  ~&=~ \left[\begin{array}{cc}
n_+ (0)_b    (14) &  0  \\
0 &  m_+ (0)_b    (14)  \\
\end{array}\right]   ~~~~~~~~~~~,~~  
{\bm {\rm L}}{}_{4}  ~
=~ \left[\begin{array}{cc}
n_+ (6)_b    (1342) &  0  \\
0 &  m_+ (6)_b    (1342)  \\
\end{array}\right]  ~~,   \cr
{\bm {\rm L}}{}_{5}  ~&=~ \left[\begin{array}{cc}
0 &  n_- (10)_b   (1243)  \\
m_- (10)_b   (1243) &   0  \\
\end{array}\right]   ~~,~~ 
{\bm {\rm L}}{}_{6}  ~=~ \left[\begin{array}{cc}
0  & n_- (12)_b    (23)  \\
m_- (12)_b    (23) &   0  \\
\end{array}\right]  ~~~~~,  \cr 
{\bm {\rm L}}{}_{7}  ~&=~ \left[\begin{array}{cc}
0 & n_- (0)_b   (14)  \\
m_- (0)_b   (14) &   0  \\
\end{array}\right]   ~~~~~~~~~~~,~~  
{\bm {\rm L}}{}_{8}  ~
=~ \left[\begin{array}{cc}
0  & n_- (6)_b    (1342)  \\
m_- (6)_b    (1342) &   0  \\
\end{array}\right]  ~~.
} \label{eq:VV}
\ee

\section{HYMN Control of `Colorful' Off-shell 4D, $\bm {\cal N} $ = 2 SUSY Multiplets}
\label{sec:HYMNC0N}

The real matrices ${\bm {\rL}}_\rI^{({\cal R})}$  and 
${\bm {\rR}}_\rI^{({\cal R})}$ 
can be used to form $16$ $\times$ $16$ matrices using the definition
\be {
{\Hat {\bm \g}}{}_\rI^{({\cal R})}  ~=~ \frac12 \, (\, {\bm  \sigma}^1 \,+\, i {\bm  \sigma}^2  \, ) \, \otimes \,
 {\bm {\rL}}_\rI^{({\cal R})}
~+~
\frac12 \, (\, {\bm  \sigma}^1 \,-\, i {\bm  \sigma}^2  \, ) \, \otimes \,
 {\bm {\rR}}_\rI^{({\cal R})}  }  ~~~.
 \label{DefGmm}
\ee
for each of the six representations,  and a corresponding matrix 
$  {\Hat {\bm {\cal C}}}^{({\cal R})}$ derived in the formula 
shown below from using each ${\Hat {\bm \g}}{}_\rI^{({\cal R})}$ representation, 
\be   
{\Hat {\bm {\cal C}}}^{({\cal R})} = {\Hat {\bm \g}}_8^{({\cal R})}   \,  \cdots   \, 
{\Hat {\bm \g}}_1^{({\cal R})}  ~~~.
\ee
This leads the way to HYMN values\cite{HYMN1,HYMN2}, which are the eigenvalues of the $ {\Hat {\bm {\cal C}}}{}^{({\cal R})}$ matrices associated with each supermultiplet. For even N, this matrix is diagonal, i.e 

$${\Hat {\bm {\cal C}}}^{({\cal R})} = \left(\begin{array}{cc}
L_{N} R_{N-1} \cdots R_{1} & 0 \\
0 & R_{N} L_{N-1} \cdots L_{1}
\end{array}\right)$$

The form of the $  {\Hat {\bm {\cal C}}}^{({\cal R})}$ matrices for each representation is shown below,
\be \eqalign{
 {\Hat {\bm {\cal C}}}{}^{(CT)} &~=~   {\Hat {\bm {\cal C}}}{}^{(CV)} ~=~{\bm  \sigma}^3\otimes  \, {\bm {\rm{I}}}{}_8   ~~~, \cr
 {\Hat {\bm {\cal C}}}{}^{(CC)} &~=~      {\Hat {\bm {\cal C}}}{}^{(TT)} ~=~ 
  {\Hat {\bm {\cal C}}}{}^{(TV)} ~=~    {\Hat {\bm {\cal C}}}{}^{(VV)}
  ~=~  \, {\bm {\rm{I}}}{}_{16}   ~~~. }
 \label{EgnV}
\ee
These results were derived by explicitly calculating all 16 $\times$ 16 matrices using
computer-enabling codes. Algorithms were written in Python and Mathematica by two independent groups. The general process of finding HYMN values is described in \cite{HYMN1} (equations 4.28-4.31), and was carried out in the same fashion by both groups, with only syntactical differences. The Python package Numpy was used extensively for matrix operations and to encode the matrices using the Numpy array type. Each group also made use of LaTeX-formatted output commands for the ease of cross-verification at intermittent steps.

The results in Eq. (\ref{EgnV}) show that the eigenvalues of $ {\Hat {\bm {\cal C}}}{}^{({\cal R})}$ split the six representations: $(CC)$, $(CT)$, $(CV)$,  $(TT)$, $(TV)$, $(VV)$, into two classes.
One class contains only $(CT)$, and $(CV)$, while the remaining four representations
are all members in a second class.

Next we calculate the anti-commutators of the matrices define in Eq. (\ref{DefGmm}).  We
find for ${({\cal R})}$ = $(CT)$, and $(CV)$
\begin{equation}
\left\{ \, {\Hat {\bm \g}}{}_\rI^{({\cal R})} ~,~ {\Hat {\bm \g}}{}_\rJ^{({\cal R})}  \, \right\} = 2\,  \delta_{\rm {IJ}} \, {\bm {\rm{I}}}{}_{16} 
~~~.
\label{CLFF2}
\end{equation}
However for the $(CC)$,  $(TT)$, $(TV)$, and $(VV)$ representations we find
\begin{equation}
\left\{ \, {\Hat {\bm \g}}{}_\rI ^{({\cal R})}~,~ {\Hat {\bm \g}}{}_\rJ^{({\cal R})}  \, \right\} = 2\,  \delta_{\rm {IJ}} \, 
{\bm {\rm{I}}}{}_{16}  ~+~ {\cal N}{}_{\rm {IJ}}{}^{\Hat \a \, ({\cal R})} \, {\bm {\kappa}}{}_{\Hat \a}^{({\cal R})}
~~~.
\label{CLFF3}
\end{equation}
where the coefficients $ {\cal N}{}_{\rm {IJ}}{}^{\Hat \a \, ({\cal R})} $ and the sets of
16 $\times$ 16 matrices ${\bm {\kappa}}{}_{\Hat \a}{}^{({\cal R})}$ are defined in equations 
(7), (30), (31), (65), (66), (74), (75), (81), and (82) of the work \cite{adnkKyeoh}
that began the line of inquiry.  We thus see an alignment between the eigenvalue classes of ${\Hat {\bm {\cal C}}}^{({\cal R})}$
and whether the  ${\bm {\rL}}_\rI{}^{({\cal R})}$ and ${\bm {\rR}}_\rI{}^{({\cal R})}$ matrices
satisfy the condition for 1D SUSY shown in Eq. (\ref{eq:GAlg1}).

The result in Eq.(\ref{CLFF3}) might 
imply that more component auxiliary fields would be needed in the cases of the 
$(CC)$, $(TT)$, $(TV)$, and $(VV)$ on-shell representations.  

Therefore, the most elegant way to understand why only the $ {({\cal R})}$ = $(CT)$, and
$(CV)$, 4D, $\cal N$ = 1 supermultiplets can describe full-fledged off-shell 4D, $\cal N$ = 2 
supermultiplets is because only their adinkras provide a spinor representation of a Euclidean 
$\mathbb {SO}(8)$ group. It is either an extraordinary coincidence that the split of the six `exemplary' on-shell supermultiplets follows the exact same ratio as the split among the six dissected
groups of ${\mathbb S}{}_4 $ \cite{pHEDRON} or there is a deeper connection yet to be uncovered.

It turns out that there is yet one more way to test the assertion
that valid off-shell supermultiplets lead to a HYMN matrix that is
traceless.  In the works of \cite{GRana1,GRana2,GHIM} which contain
our earliest exploration of these issues, an algorithm is given for constructing L-matrices that is {\it {independent}}
of combination of $\cal N$ = 1 supermultiplets that can be combined into a valid $\cal N$ = 2 supermultiplet.  One set of these L-matrices leads to an octet that contains the identity matrix and
can be dubbed the `Diadem(8)' Octet (denoted by ${\cal D}$O(8)). This is the higher dimensional analogue\footnote{In the sense that its matrices are 8 $\times$ 8 matrices.} of the Klein 4-group (also known as the Klein Vierergruppe). From the work in \cite{GRana1} the elements of this octet are given by

$$\begin{aligned}
&\mathbf{L}_{1}=\mathbf{I}_{2 \times 2} \otimes \mathbf{I}_{2 \times 2} \otimes \mathbf{I}_{2 \times 2}=\mathbf{R}_{1}\\
&\mathbf{L}_{2}=i \mathbf{I}_{2 \times 2} \otimes \boldsymbol{\sigma}^{3} \otimes \boldsymbol{\sigma}^{2}=-\mathbf{R}_{2}\\
&\mathbf{L}_{3}=i \boldsymbol{\sigma}^{3} \otimes \boldsymbol{\sigma}^{2} \otimes \mathbf{I}_{2 \times 2}=-\mathbf{R}_{3}\\
&\mathbf{L}_{4}=i \mathbf{I}_{2 \times 2} \otimes \boldsymbol{\sigma}^{1} \otimes \boldsymbol{\sigma}^{2}=-\mathbf{R}_{4}\\
&\mathbf{L}_{5}=i \boldsymbol{\sigma}^{1} \otimes \boldsymbol{\sigma}^{2} \otimes \mathbf{I}_{2 \times 2}=-\mathbf{R}_{5}\\
&\mathbf{L}_{6}=i \boldsymbol{\sigma}^{2} \otimes \mathbf{I}_{2 \times 2} \otimes \boldsymbol{\sigma}^{1}=-\mathbf{R}_{6}\\
&\mathbf{L}_{7}=i \boldsymbol{\sigma}^{2} \otimes \mathbf{I}_{2 \times 2} \otimes \boldsymbol{\sigma}^{3}=-\mathbf{R}_{7}\\
&\mathbf{L}_{8}=i \boldsymbol{\sigma}^{2} \otimes \boldsymbol{\sigma}^{2} \otimes \boldsymbol{\sigma}^{2}=-\mathbf{R}_{8}
\end{aligned}$$
and it leads to the result
\be   
{\Hat {\bm {\cal C}}}^{({\cal D}{\rm O}(8))} = {\Hat {\bm \g}}_8^{({\cal D}{\rm O}(8))}   \,  \cdots   \, 
{\Hat {\bm \g}}_1^{(({\cal D}{\rm O}(8)))} ~=~{\bm  \sigma}^3\otimes  \, {\bm {\rm{I}}}{}_8   ~~~, 
\ee
showing it is indeed in the same class that also contains $(CT)$ and $(CV)$.

\section{4D, $\cal N$ = 1 SUSY and the Permutahedron}
\label{sec:RecRev1}

\begin{figure}[h]
\includegraphics[width=13cm]{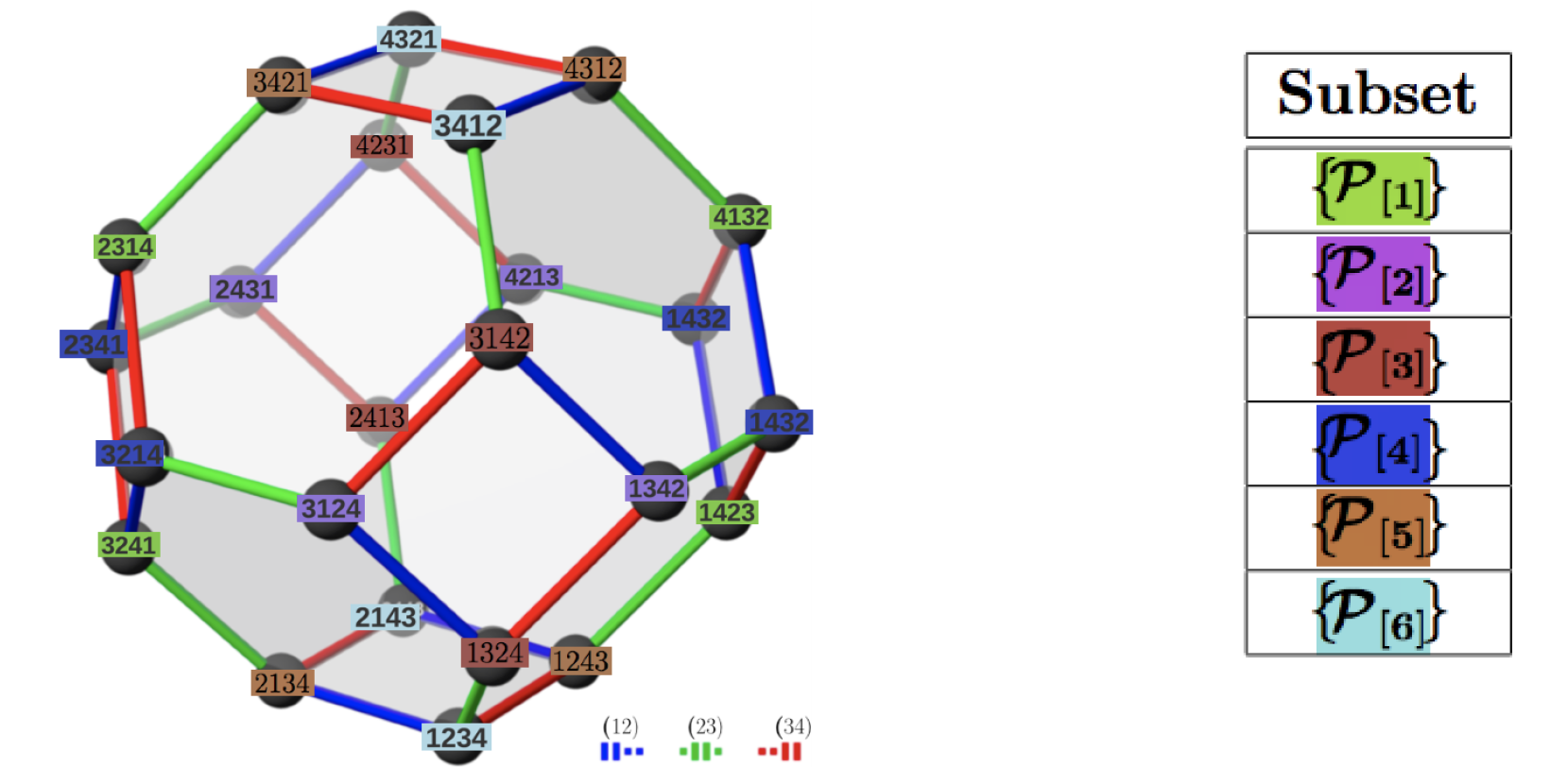}
\centering
\caption{Colored permutation addresses adorning the permutahedron}
\label{perm}
\end{figure}

In a recent paper, \cite{pHEDRON} the relevance of a well-known mathematical concept, the 
permutahedron\cite{pHR0n1,pHR0n2,pHR0n3,pHR0n4}, was brought into focus.  In particular, 
it was conjectured the permutahedron for ${\mathbb S}{}_4 $, along with Bruhat weak ordering 
\cite{BruHT}, provide a foundation for a representation theory of off-shell 4D, 
$\cal N$ = 1 SUSY theories.  A representation of the ${\mathbb S}{}_4 $ permutahedron is shown 
in Fig. \ref{perm}.  Each of the listed subsets (dubbed quartets) contains four elements. Those four elements are shown in the same color on the permutahedron.

We remind the reader that of the subsets, only the $\{  {\cal P}{}_{[1]} \}$, $\{  {\cal P}{}_{[2]} \}$, and
$\{  {\cal P}{}_{[3]} \}$ ones are associated, respectively, with the chiral, tensor, and vector supermultiplets.  
These were derived by a projection procedure carried out in the work of \cite{permutadnk}. Respectively, these are associated with the green, purple and rust colored permutation elements.

There is one special property of $\{  {\cal P}{}_{[6]} \}$ as it is the only subset that contains the identity permutation element. Regarding all the subsets as unordered, the following equations follow,
$$
\{ {\cal P}{}_{[1]}\} ~=~ (132) \, \{ {\cal P}{}_{[6]}\} ~~~,~~~
\{ {\cal P}{}_{[2]}\} ~=~ (123) \, \{ {\cal P}{}_{[6]}\}  ~~~,~~~
\{ {\cal P}{}_{[4]}\} ~=~ (23) \, \{ {\cal P}{}_{[6]}\}  ~~~,
$$
\be
\{ {\cal P}{}_{[4]}\} ~=~ (13) \, \{ {\cal P}{}_{[6]}\} ~~~,~~~
\{ {\cal P}{}_{[5]}\} ~=~ (12) \, \{ {\cal P}{}_{[6]}\}  ~~~,~~~
\ee
and according to the use of a lexicographical ordering prescription $\{  {\cal P}{}_{[6]} \}$ is the 
``smallest'' of the twenty-four permutations contained in ${\mathbb S}_4 $. The permutations in the subset $ \{ {\cal P}{}_{[6]}\}$ can also be represented as 
\be \begin{array}{ccccccccc}
{\bm {\cal P}}_{1} &=&  \,{\bm {\rm I}}{}_{2} & \otimes &{\bm {\rm I}}
{}_{2} & = & () & = & \langle 1 2 3 4 \rangle  ~~~, \\
{\bm {\cal P}}_{2} &=&{\bm {\rm I}}{}_{2}  & \otimes &{\bm \s}^1  & =&  (12)(34)  &= &  \langle 2 1  4 3  \rangle   ~~~, \\ 
{\bm {\cal P}}_{3} &=& {\bm \s}^1 & \otimes &{\bm {\rm I}}{}_{2}  & =&  (13)(24) & = &  \langle 3 4 1 2 \rangle ~~~. \\
{\bm {\cal P}}_{4} &=& {\bm \s}^1 & \otimes &{\bm \s}^1  &=& (14)(23)  & = &   \langle 4 3 2 1 \rangle
~~~, \\
\end{array} \label{N4dia}  \ee
written successively in matrix, cycle, and one-line notations.

The permutations in Eq.\ ({\ref{N4dia}}), which correspond to the supercharges that act on the ${{\cal P}}_{6}$ supermultiplet, occur at two vertices of the base face and two vertices of the top face. In this sense, they can be seen as two-colored quartets.  Further explanation of this statement is warranted, for which we turn to the findings of \cite{GRana1}, in particular appendix A. Given an arbitrary $\cal N$-extended theory with N matrices, we can always build an ($\cal N$-1)-extended theory by taking any set of (N-1) matrices from the $\cal N$-extended theory. For the case of 4D, $\cal N$ = 1 SUSY, after dimensional reduction we end up with a 1D, $\cal N$ = 4 theory \cite{holography}. The process, which shows the “interdimensionality” of adinkras, is as follows. Reducing to 3D, we get an $\cal N$ = 2 scalar theory (chirality doesn’t exist in 3D). Further reducing to 2D, we get a 2D, $\cal N$ = (1, 1) theory. Reduction once more leads to the 1D, $\cal N$ = 4 scalar theory. All of these theories have the same adinkras. Turning back to the present case of the 1D, $\cal N$ = 4 theory, it follows that we could use this $\cal N$ = 4 theory to build theories for $\cal N$ = 1, 2, or 3, by simply taking subsets of the quartets. In this particular way, the faces of the permutahedron for the 1D, $\cal N$ = 4 theory are seen to be composed of the supercharges that act on two different $\cal N$ = 2 supermultiplets (where each $\cal N$ = 2 supermultiplet is acted on by 2 of the matrices corresponding to the supercharges for the larger $\cal N$ = 4 supermultiplets). 

A similar argument is constructed in \cite{Toppan}, where the actions for 1D, $\cal N$ = 4 $\sigma$ -models are computed with respect to the two inequivalent (2,8,6) multiplets. It's shown that imposing the 5th supersymmetry (with four supersymmetry generators already manifest) automatically induces full $\cal N$ = 8 off-shell invariance.

This has an interesting implication for the 4D, ${\cal N} $ = 2 theories that are the target of our investigation in this work. In the future we seek to further understand the role that the permutahedron may play in picking out valid combinations of $\cal N$ = 1 supermultiplets to create valid $\cal N$ = 2 supermultiplets. The fact that permutations corresponding to the supercharges of valid supermultiplets of lower degree or extension may be formed by the square faces hints at a potential way to embed lower degree supercharges of supermultiplets (say of $\cal N$ = 1) into higher dimensional permutahedra (say of $\cal N$ = 2). This will be investigated in future works.


\section{ `Colorful' Off-shell 4D, $\bm {\cal N} $ = 2 SUSY Multiplets \& Their Explicit
{$\mathbb S{}_8$} Permutations}
\label{sec:S8perm}

The skeptical reader may object that arguments based on the use of $ {\Hat {\bm \g}}{}_\rI{}^{({\cal R})}$ 
having nothing to do with the permutahedron associated with
${\mathbb S}{}_8$.  This is not so.  As indicated by the result in Eq.\ (\ref{eq:aas1}),
the permutation elements powerfully determine the forms of the matrices that are
described by the ${\cal {GR}}(d, N)$ or Garden Algebra in Eq.\ (\ref{eq:GAlg1}).
We believe this current work shows how the ${\cal {GR}}$(8, $8)$ representation
clearly impacts how off-shell 4D, $\cal N$ = 1 theories can be combined to become
off-shell 4D, $\cal N$ = 2 theories. 

As a step toward enabling a deeper study of these issues, it is necessary to give a reformulation of the results in (\ref{eq:CC}) -  (\ref{eq:VV}).  We find these can be recast as the following, where explicit dependences on the elements in ${\mathbb S}{}_8$ (as both cycle and one-line notations) appear.

${ ({\cal R}) \,=\, (\bm{CC})}$
\be
\begin{array}{cccccc}
~{\bm {\rL}}_1 & ~=~ &  ~~(170)_b(243)(687) & ~=~ & ~~  (170)_b
\langle \, 1 \, 4 \,  2 \,  3 \,  5 \,  8 \,  6 \,  7 \,  \rangle     &~~~,  \\
~{\bm {\rL}}_2 & ~=~ &  ~~(204)_b(123)(567)      & ~=~ & ~~(204)_b
\langle \,   2 \,  3 \, 1 \, 4 \, 6 \, 7 \, 5 \,  8 \, \rangle         &~~~,  \\
~{\bm {\rL}}_3 & ~=~ &  ~~(102)_b(134)(578)     & ~=~ & ~~(102)_b
\langle \,  3 \,  2 \,  4 \,  1 \,  7 \,  6 \,  8 \,  5     \rangle    & ~~~,  \\
~{\bm {\rL}}_4 & ~=~ &  ~~(0)_b(142)(586)     & ~=~ & ~~(0)_b
\langle \,   4 \,  1 \,  3 \,  2 \, 8 \,  5 \,  7 \,  6 \, \rangle    &~~~,  \\
~{\bm {\rL}}_5 & ~=~ &  ~~(15)_b(15)(364728)     & ~=~ & ~~(15)_b
\langle \, 5 \, 8 \, 6 \, 7 \, 1 \,  4 \,  2 \, 3  \, \rangle     &~~~,  \\
~{\bm {\rL}}_6 & ~=~ &  ~~(105)_b(163527)(48)      & ~=~ & ~~(105)_b 
\langle \, 6 \, 7 \,  5 \,  8 \,  2 \,  3 \,  1 \,  4 \,   \rangle    &~~~,  \\
~{\bm {\rL}}_7 & ~=~ &  ~~(195)_b(26)(174538)      & ~=~ & ~~ (195)_b
\langle \,  7 \, 6 \, 8 \, 5 \, 3 \,  2 \,  4 \,  1  \,  \rangle     &~~~,  \\
~{\bm {\rL}}_8 & ~=~ &  ~~(165)_b(37)(182546)     & ~=~ & ~~ (165)_b 
\langle \, 8 \, 5 \, 7 \, 6 \, 4 \, 1 \, 3 \, 2 \, \rangle   &~~~,  \\
\end{array}
\label{cmcm}
\ee

${ ({\cal R}) \,=\, (\bm{CT})}$
$$ 
\begin{array}{cccccc}
~~~~~~~{\bm {\rL}}_1 & ~=~ &  ~~~(234)_b(243)(678) {~~~}&~~~~~~=~ & ~~  (234)_b
\langle \, 1 \, 4 \,  2 \,  3 \,  5 \,  7 \,  8 \,  6 \,  \rangle &\,~\,~, ~\,~ {\,} \\
~~~~~~~{\bm {\rL}}_2 & ~=~ &  (76)_b(123)(568)     &~~~~~~=~ & ~~ (76)_b
\langle \,   2 \,  3 \, 1 \, 4 \, 6 \, 8 \, 7 \,  5 \, \rangle  &,   \\
~~~~~~~{\bm {\rL}}_3 & ~=~ &  (134)_b(134)(576)     &~~~~~~=~ & ~~ (134)_b
\langle \,  3 \,  2 \,  4 \,  1 \,  7 \,  5 \,  6 \,  8     \rangle  &, \\
~~~~~~~{\bm {\rL}}_4 & ~=~ &  (32)_b(142)(587)     &~~~~~~=~ & ~~ (32)_b
\langle \,   4 \,  1 \,  3 \,  2 \, 8 \,  6 \,  5 \,  7 \, \rangle  &,  \\
\end{array}
$$

\be
\begin{array}{cccccc}
~~~~{\bm {\rL}}_5~~~ & ~=~ &  ~~(11)_b(15)(28)(36)(47)      & ~=~ & ~~ (11)_b
\langle \, 5 \, 8 \, 6 \, 7 \, 1 \,  3 \,  4 \, 2  \, \rangle  &~~~,  \\
~~~~{\bm {\rL}}_6~~~ & ~=~ &  ~~(173)_b(1648)(2735)     & ~=~ & ~~ (173)_b 
\langle \, 6 \, 7 \,  5 \,  8 \,  2 \,  4 \,  3 \,  1 \,   \rangle  &~~~,   \\
~~~~{\bm {\rL}}_7~~~ & ~=~ &  ~~ (103)_b(1726)(3845)    & ~=~ & ~~ (103)_b
\langle \,  7 \, 6 \, 8 \, 5 \, 3 \,  1 \,  2 \,  4  \,  \rangle  &~~~,   \\
~~~~{\bm {\rL}}_8~~~ & ~=~ &  ~~(193)_b(1837)(2546)     & ~=~ & ~~  (193)_b 
\langle \, 8 \, 5 \, 7 \, 6 \, 4 \, 2 \, 1 \, 3 \, \rangle  &~~~,  \\
\end{array}
\ee

${ ({\cal R}) \,=\, (\bm{CV})}$
\be
\begin{array}{cccccc}
~{\bm {\rL}}_1 & ~=~ &  ~~(170)_b (243)(5687) & ~=~ & ~~  (170)_b
\langle \, 1 \, 4 \,  2 \,  3 \,  6 \,  8 \,  5 \,  7 \,  \rangle  &~~~,   \\
~{\bm {\rL}}_2 & ~=~ &  ~~(204)_b(123)(67)     & ~=~ & ~~ (204)_b
\langle \,   2 \,  3 \, 1 \, 4 \, 5 \, 7 \, 6 \,  8 \, \rangle  &~~~,  \\
~{\bm {\rL}}_3 & ~=~ &  ~~(6)_b(134)(58)     & ~=~ & ~~ (6)_b
\langle \,  3 \,  2 \,  4 \,  1 \,  8 \,  6 \,  7 \,  5     \rangle  &~~~, \\
~{\bm {\rL}}_4 & ~=~ &  ~~(96)_b(142)(5786)     & ~=~ & ~~ (96)_b
\langle \,   4 \,  1 \,  3 \,  2 \, 7 \,  5 \,  8 \,  6 \, \rangle  &~~~,  \\
~{\bm {\rL}}_5 & ~=~ &  ~~(210)_b(15283647)      & ~=~ & ~~ (210)_b
\langle \, 5 \, 8 \, 6 \, 7 \, 2 \,  4 \,  1 \, 3  \, \rangle  &~~~,   \\
~{\bm {\rL}}_6 & ~=~ &  ~~(180)_b(1635)(27)(48)      & ~=~ & ~~(180)_b 
\langle \, 6 \, 7 \,  5 \,  8 \,  1 \,  3 \,  2 \,  4 \,   \rangle  &~~~,  \\
~{\bm {\rL}}_7 & ~=~ &  ~~(126)_b(1738)(26)(45)      & ~=~ & ~~(126)_b
\langle \,  7 \, 6 \, 8 \, 5 \, 4 \,  2 \,  3 \,  1  \,  \rangle  &~~~,   \\
~{\bm {\rL}}_8 & ~=~ &  ~~(24)_b(18253746)     & ~=~ & ~~(24)_b 
\langle \, 8 \, 5 \, 7 \, 6 \, 3 \, 1 \, 4 \, 2 \, \rangle  &~~~, \\
\end{array}
\ee

${ ({\cal R}) \,=\, (\bm{TT})}$
$$ \eqalign{
{\bm {\rL}}_1 ~&~=~ (238)_b \, \left[ \, \left( \fracm12  \right)  \left( \, ({\bm {\rI}} + {\bm {\s^3}})n_+ + \,({\bm {\rI}} - {\bm {\s^3}})m_+ \, \right) \, \otimes \, {{\bm {\rI}}}_4 \, \right] (234)(678) \cr ~&~=~ (238)_b \, \left[ \,  \left( \fracm12  \right) \left( \, \,({\bm {\rI}} + {\bm {\s^3}})n_+ + \,({\bm {\rI}} - {\bm {\s^3}})m_+ \, \right) \, \otimes \, {{\bm {\rI}}}_4 \, \right]
\langle \, 1 \, 3 \,  4 \,  2 \,  5 \,  7 \,  8 \,  6 \,   \rangle    ~~, \cr
{\bm {\rL}}_2 ~&~=~(68)_b \, \left[ \, \left( \fracm12  \right)  \left( \, ({\bm {\rI}} + {\bm {\s^3}})n_+ + \,({\bm {\rI}} - {\bm {\s^3}})m_+ \, \right) \, \otimes \, {{\bm {\rI}}}_4 \, \right] (124)(568) \cr ~&~=~ (68)_b \, \left[ \,  \left( \fracm12  \right) \left( \, \,({\bm {\rI}} + {\bm {\s^3}})n_+ + \,({\bm {\rI}} - {\bm {\s^3}})m_+ \, \right) \, \otimes \, {{\bm {\rI}}}_4 \, \right]
\langle \,   2 \,  4 \, 3 \, 1 \, 6 \, 8 \, 7 \,  5 \,  \rangle    \,~~~,  \cr
{\bm {\rL}}_3 ~&~=~(136)_b \, \left[ \,  \left( \fracm12  \right) \left( \, ({\bm {\rI}} + {\bm {\s^3}})n_+ + \,({\bm {\rI}} - {\bm {\s^3}})m_+ \, \right) \, \otimes \, {{\bm {\rI}}}_4 \, \right] (132)(576) \cr ~&~=~ (136)_b \, \left[ \,  \left( \fracm12  \right) \left( \, \,({\bm {\rI}} + {\bm {\s^3}})n_+ + \,({\bm {\rI}} - {\bm {\s^3}})m_+ \, \right) \, \otimes \, {{\bm {\rI}}}_4 \, \right]
\langle \,  3 \,  1 \,  2 \,  4 \,  7 \,  5 \,  6 \,  8      \rangle    ~~~,\cr
{\bm {\rL}}_4 ~&~=~(34)_b \, \left[ \,  \left( \fracm12  \right)  \left( \, ({\bm {\rI}} + {\bm {\s^3}})n_+ + \,({\bm {\rI}} - {\bm {\s^3}})m_+ \, \right) \, \otimes \, {{\bm {\rI}}}_4 \, \right] (143)(587) \cr ~&~=~ (34)_b \, \left[ \,  \left( \fracm12  \right)  \left( \, ({\bm {\rI}} + {\bm {\s^3}})n_+ + \,({\bm {\rI}} - {\bm {\s^3}})m_+ \, \right) \, \otimes \, {{\bm {\rI}}}_4 \, \right]
\langle \,   4 \,  2 \,  1 \,  3 \, 8 \,  6 \,  5 \,  7 \,  \rangle    ~~~~, \cr
{\bm {\rL}}_5 ~&~=~(238)_b \, \left[ \,  \left( \fracm12  \right)  \left( \, ({\bm {\rI}} + {\bm {\s^3}})n_- + \,({\bm {\rI}} - {\bm {\s^3}})m_- \, \right) \, \otimes \, {{\bm {\rI}}}_4 \, \right] (15)(274638) \cr ~&~=~ (238)_b \, \left[ \,  \left( \fracm12  \right)  \left( \, \,({\bm {\rI}} + {\bm {\s^3}})n_- + \,({\bm {\rI}} - {\bm {\s^3}})m_- \, \right) \, \otimes \, {{\bm {\rI}}}_4 \, \right]
\langle \, 5 \, 7 \, 8 \, 6 \, 1 \,  3 \,  4 \, 2  \,  \rangle    ~~,  \cr
{\bm {\rL}}_6 ~&~=~(68)_b \, \left[ \,  \left( \fracm12  \right) \left(  \,({\bm {\rI}} + {\bm {\s^3}})n_- + \,({\bm {\rI}} - {\bm {\s^3}})m_- \, \right) \, \otimes \, {{\bm {\rI}}}_4 \, \right] (164528)(37) \cr ~&~=~ (68)_b \, \left[ \,  \left( \fracm12  \right) \left( \, \,({\bm {\rI}} + {\bm {\s^3}})n_- + \,({\bm {\rI}} - {\bm {\s^3}})m_- \, \right) \, \otimes \, {{\bm {\rI}}}_4 \, \right]
\langle \, 6 \, 8 \,  7 \,  5 \, 2 \,  4 \,  3 \,  1 \,    \rangle    \,~~~,  \cr
{\bm {\rL}}_7 ~&~=~(136)_b \, \left[ \,  \left( \fracm12  \right)  \left(  \,({\bm {\rI}} + {\bm {\s^3}})n_- + \,({\bm {\rI}} - {\bm {\s^3}})m_- \, \right) \, \otimes \, {{\bm {\rI}}}_4 \, \right] (172536)(48) \cr ~&~=~ (136)_b \, \left[ \,  \left( \fracm12  \right) \left( \, ({\bm {\rI}} + {\bm {\s^3}})n_- + \,({\bm {\rI}} - {\bm {\s^3}})m_- \, \right) \, \otimes \, {{\bm {\rI}}}_4 \, \right]
\langle \,  7 \, 5 \, 6 \, 8 \, 3 \,  1 \,  2 \,  4  \,   \rangle    ~\,~,  
}
$$ 
\be \eqalign{
{\bm {\rL}}_8 ~&~\,=~(34)_b \, \left[ \,  \left( \fracm12  \right) \left( \, ({\bm {\rI}} + {\bm {\s^3}})n_- + \,({\bm {\rI}} - {\bm {\s^3}})m_- \, \right) \, \otimes \, {{\bm {\rI}}}_4 \, \right] (183547)(26) \cr ~&\,~=~ (34)_b \, \left[ \,  \left( \fracm12  \right) \left( \, ({\bm {\rI}} + {\bm {\s^3}})n_- + \,({\bm {\rI}} - {\bm {\s^3}})m_- \, \right) \, \otimes \, {{\bm {\rI}}}_4 \, \right]
\langle \, 8 \, 6 \, 5 \, 7 \, 4 \, 2 \, 1 \, 3 \,  \rangle    ~~~~, \cr
}
\ee

${ ({\cal R}) \,=\, (\bm{TV})}$
\be \eqalign{
{\bm {\rL}}_1 ~&~=~ (174)_b \, \left[ \, \left( \fracm12  \right) \, \left( \, \,({\bm {\rI}} + {\bm {\s^3}})n_+ + \,({\bm {\rI}} - {\bm {\s^3}})m_+ \, \right) \, \otimes \, {{\bm {\rI}}}_4 \, \right] (234)(5687) \cr ~&~=~ (174)_b \, \left[ \, \left( \fracm12  \right) \, \left( \, \,({\bm {\rI}} + {\bm {\s^3}})n_+ + \,({\bm {\rI}} - {\bm {\s^3}})m_+ \, \right) \, \otimes \, {{\bm {\rI}}}_4 \, \right]
\langle \, 1 \, 3 \,  4 \,  2 \,  6 \,  8 \,  5 \,  7 \,   \rangle    ~~, \cr
{\bm {\rL}}_2 ~&~=~(196)_b \, \left[ \, \left( \fracm12  \right) \, \left( \, \,({\bm {\rI}} + {\bm {\s^3}})n_+ + \,({\bm {\rI}} - {\bm {\s^3}})m_+ \, \right) \, \otimes \, {{\bm {\rI}}}_4 \, \right] (124)(67) \cr ~&~=~ (196)_b \, \left[ \, \left( \fracm12  \right) \, \left( \, \,({\bm {\rI}} + {\bm {\s^3}})n_+ + \,({\bm {\rI}} - {\bm {\s^3}})m_+ \, \right) \, \otimes \, {{\bm {\rI}}}_4 \, \right]
\langle \,   2 \,  4 \, 3 \, 1 \, 5 \, 7 \, 6 \,  8 \,  \rangle    \,~~~,  \cr
{\bm {\rL}}_3 ~&~=~(8)_b \, \left[ \, \left( \fracm12  \right) \, \left( \, \,({\bm {\rI}} + {\bm {\s^3}})n_+ + \,({\bm {\rI}} - {\bm {\s^3}})m_+ \, \right) \, \otimes \, {{\bm {\rI}}}_4 \, \right] (132)(58) \cr ~&~=~ (8)_b \, \left[ \, \left( \fracm12  \right) \, \left( \, \,({\bm {\rI}} + {\bm {\s^3}})n_+ + \,({\bm {\rI}} - {\bm {\s^3}})m_+ \, \right) \, \otimes \, {{\bm {\rI}}}_4 \, \right]
\langle \,  3 \,  1 \,  2 \,  4 \,  8 \,  6 \,  7 \,  5      \rangle    \,~~,\cr
{\bm {\rL}}_4 ~&~=~(98)_b \, \left[ \, \left( \fracm12  \right) \, \left( \, \,({\bm {\rI}} + {\bm {\s^3}})n_+ + \,({\bm {\rI}} - {\bm {\s^3}})m_+ \, \right) \, \otimes \, {{\bm {\rI}}}_4 \, \right] (143)(5786) \cr ~&~=~ (98)_b \, \left[ \, \left( \fracm12  \right) \, \left( \, \,({\bm {\rI}} + {\bm {\s^3}})n_+ + \,({\bm {\rI}} - {\bm {\s^3}})m_+ \, \right) \, \otimes \, {{\bm {\rI}}}_4 \, \right]
\langle \,   4 \,  2 \,  1 \,  3 \, 7 \,  5 \,  8 \,  6 \,  \rangle    \,~~~, \cr
{\bm {\rL}}_5 ~&~=~(174)_b \, \left[ \, \left( \fracm12  \right) \, \left( \, \,({\bm {\rI}} + {\bm {\s^3}})n_- + \,({\bm {\rI}} - {\bm {\s^3}})m_- \, \right) \, \otimes \, {{\bm {\rI}}}_4 \, \right] (38)(46)(1527) \cr ~&~=~ (174)_b \, \left[ \, \left( \fracm12  \right) \, \left( \, \,({\bm {\rI}} + {\bm {\s^3}})n_- + \,({\bm {\rI}} - {\bm {\s^3}})m_- \, \right) \, \otimes \, {{\bm {\rI}}}_4 \, \right]
\langle \, 5 \, 7 \, 8 \, 6 \, 2 \,  4 \,  1 \, 3  \,  \rangle    ~~,  \cr
{\bm {\rL}}_6 ~&~=~(196)_b \, \left[ \, \left( \fracm12  \right) \, \left( \, \,({\bm {\rI}} + {\bm {\s^3}})n_- + \,({\bm {\rI}} - {\bm {\s^3}})m_- \, \right) \, \otimes \, {{\bm {\rI}}}_4 \, \right] (16372845) \cr ~&~=~ (196)_b \, \left[ \, \left( \fracm12  \right) \, \left( \, \,({\bm {\rI}} + {\bm {\s^3}})n_- + \,({\bm {\rI}} - {\bm {\s^3}})m_- \, \right) \, \otimes \, {{\bm {\rI}}}_4 \, \right]
\langle \, 6 \, 8 \,  7 \,  5 \, 1 \,  3 \,  2 \,  4 \,    \rangle    \,~~~,  \cr
{\bm {\rL}}_7 ~&~=~(8)_b \, \left[ \, \left( \fracm12  \right) \, \left( \, \,({\bm {\rI}} + {\bm {\s^3}})n_- + \,({\bm {\rI}} - {\bm {\s^3}})m_- \, \right) \, \otimes \, {{\bm {\rI}}}_4 \, \right] (17362548) \cr ~&~=~ (8)_b \, \left[ \, \left( \fracm12  \right) \, \left( \, \,({\bm {\rI}} + {\bm {\s^3}})n_- + \,({\bm {\rI}} - {\bm {\s^3}})m_- \, \right) \, \otimes \, {{\bm {\rI}}}_4 \, \right]
\langle \,  7 \, 5 \, 6 \, 8 \, 4 \,  2 \,  3 \,  1  \,   \rangle    \,~\, ,  \cr
{\bm {\rL}}_8 ~&~=~(98)_b \, \left[ \, \left( \fracm12  \right) \, \left( \, \,({\bm {\rI}} + {\bm {\s^3}})n_- + \,({\bm {\rI}} - {\bm {\s^3}})m_- \, \right) \, \otimes \, {{\bm {\rI}}}_4 \, \right] (35)(47)(1862) \cr ~&~=~ (98)_b \, \left[ \, \left( \fracm12  \right) \, \left( \, \,({\bm {\rI}} + {\bm {\s^3}})n_- + \,({\bm {\rI}} - {\bm {\s^3}})m_- \, \right) \, \otimes \, {{\bm {\rI}}}_4 \, \right]
\langle \, 8 \, 6 \, 5 \, 7 \, 3 \, 1 \, 4 \, 2 \,  \rangle    ~~\,~, \cr
}
\ee

${ ({\cal R}) \,=\, (\bm{VV})}$
$$ \eqalign{
{\bm {\rL}}_1 ~&~=~ (170)_b \, \left[ \, \left( \fracm12  \right) \, \left( \, \,({\bm {\rI}} + {\bm {\s^3}})n_+ + \,({\bm {\rI}} - {\bm {\s^3}})m_+ \, \right) \, \otimes \, {{\bm {\rI}}}_4 \, \right] (1243)(5687) \cr ~&~=~ (170)_b \, \left[ \, \left( \fracm12  \right) \, \left( \, \,({\bm {\rI}} + {\bm {\s^3}})n_+ + \,({\bm {\rI}} - {\bm {\s^3}})m_+ \, \right) \, \otimes \, {{\bm {\rI}}}_4 \, \right]
\langle \, 2 \, 4 \,  1 \,  3 \,  6 \,  8 \,  5 \,  7 \,   \rangle    ~~~~~~~, \cr
{\bm {\rL}}_2 ~&~=~(204)_b \, \left[ \, \left( \fracm12  \right) \, \left( \, \,({\bm {\rI}} + {\bm {\s^3}})n_+ + \,({\bm {\rI}} - {\bm {\s^3}})m_+ \, \right) \, \otimes \, {{\bm {\rI}}}_4 \, \right] (23)(67) \cr ~&~=~ (204)_b \, \left[ \, \left( \fracm12  \right) \, \left( \, \,({\bm {\rI}} + {\bm {\s^3}})n_+ + \,({\bm {\rI}} - {\bm {\s^3}})m_+ \, \right) \, \otimes \, {{\bm {\rI}}}_4 \, \right]
\langle \,   1 \,  3 \, 2 \, 4 \, 5 \, 7 \, 6 \,  8 \,  \rangle    ~~~~~~~,  \cr
{\bm {\rL}}_3 ~&~=~(0)_b \, \left[ \, \left( \fracm12  \right) \, \left( \, \,({\bm {\rI}} + {\bm {\s^3}})n_+ + \,({\bm {\rI}} - {\bm {\s^3}})m_+ \, \right) \, \otimes \, {{\bm {\rI}}}_4 \, \right] (14)(58) \cr ~&~=~ (0)_b \, \left[ \, \left( \fracm12  \right) \, \left( \, \,({\bm {\rI}} + {\bm {\s^3}})n_+ + \,({\bm {\rI}} - {\bm {\s^3}})m_+ \, \right) \, \otimes \, {{\bm {\rI}}}_4 \, \right]
\langle \,  4 \,  2 \,  3 \,  1 \,  8 \,  6 \,  7 \,  5      \rangle    ~~~~~~~~~~,
}
$$

\be \eqalign{
{\bm {\rL}}_4 ~&~=~(102)_b \, \left[ \, \left( \fracm12  \right) \, \left( \, \,({\bm {\rI}} + {\bm {\s^3}})n_+ + \,({\bm {\rI}} - {\bm {\s^3}})m_+ \, \right) \, \otimes \, {{\bm {\rI}}}_4 \, \right] (1342)(5786) \cr ~&~=~ (102)_b \, \left[ \, \left( \fracm12  \right) \, \left( \, \,({\bm {\rI}} + {\bm {\s^3}})n_+ + \,({\bm {\rI}} - {\bm {\s^3}})m_+ \, \right) \, \otimes \, {{\bm {\rI}}}_4 \, \right]
\langle \,   3 \,  1 \,  4 \,  2 \, 7 \,  5 \,  8 \,  6 \,  \rangle    ~~~~~~,  \cr
{\bm {\rL}}_5 ~&~=~(170)_b \, \left[ \, \left( \fracm12  \right) \, \left( \, \,({\bm {\rI}} + {\bm {\s^3}})n_- + \,({\bm {\rI}} - {\bm {\s^3}})m_- \, \right) \, \otimes \, {{\bm {\rI}}}_4 \, \right] (1647)(2835) \cr ~&~=~ (170)_b \, \left[ \, \left( \fracm12  \right) \, \left( \, \,({\bm {\rI}} + {\bm {\s^3}})n_- + \,({\bm {\rI}} - {\bm {\s^3}})m_- \, \right) \, \otimes \, {{\bm {\rI}}}_4 \, \right]
\langle \, 6 \, 8 \, 5 \, 7 \, 2 \,  4 \,  1 \, 3  \,  \rangle    ~~~~~~,  \cr
{\bm {\rL}}_6 ~&~=~(204)_b \, \left[ \, \left( \fracm12  \right) \, \left( \, \,({\bm {\rI}} + {\bm {\s^3}})n_- + \,({\bm {\rI}} - {\bm {\s^3}})m_- \, \right) \, \otimes \, {{\bm {\rI}}}_4 \, \right] (15)(27)(36)(48) \cr ~&~=~ (204)_b \, \left[ \, \left( \fracm12  \right) \, \left( \, \,({\bm {\rI}} + {\bm {\s^3}})n_- + \,({\bm {\rI}} - {\bm {\s^3}})m_- \, \right) \, \otimes \, {{\bm {\rI}}}_4 \, \right]
\langle \, 5 \, 7 \,  6 \,  8 \, 1 \,  3 \,  2 \,  4 \,    \rangle    ~~~~~~~,  \cr
{\bm {\rL}}_7 ~&~=~(0)_b \, \left[ \, \left( \fracm12  \right) \, \left( \, \,({\bm {\rI}} + {\bm {\s^3}})n_- + \,({\bm {\rI}} - {\bm {\s^3}})m_- \, \right) \, \otimes \, {{\bm {\rI}}}_4 \, \right] (18)(26)(37)(45) \cr ~&~=~ (0)_b \, \left[ \, \left( \fracm12  \right) \, \left( \, \,({\bm {\rI}} + {\bm {\s^3}})n_- + \,({\bm {\rI}} - {\bm {\s^3}})m_- \, \right) \, \otimes \, {{\bm {\rI}}}_4 \, \right]
\langle \,  8 \, 6 \, 7 \, 5 \, 4 \,  2 \,  3 \,  1  \,   \rangle    ~~~~~~~~~~,  \cr
{\bm {\rL}}_8 ~&~\,=~(102)_b \, \left[ \, \left( \fracm12  \right) \, \left( \, \,({\bm {\rI}} + {\bm {\s^3}})n_- + \,({\bm {\rI}} - {\bm {\s^3}})m_- \, \right) \, \otimes \, {{\bm {\rI}}}_4 \, \right] (1746)(2538) \cr ~&~=~ (102)_b \, \left[ \, \left( \fracm12  \right) \, \left( \, \,({\bm {\rI}} + {\bm {\s^3}})n_- + \,({\bm {\rI}} - {\bm {\s^3}})m_- \, \right) \, \otimes \, {{\bm {\rI}}}_4 \, \right]
\langle \, 7 \, 5 \, 8 \, 6 \, 3 \, 1 \, 4 \, 2 \,  \rangle    ~~~~~~~. \cr
}
\ee

\begin{figure}[h]
\includegraphics[width=8cm]{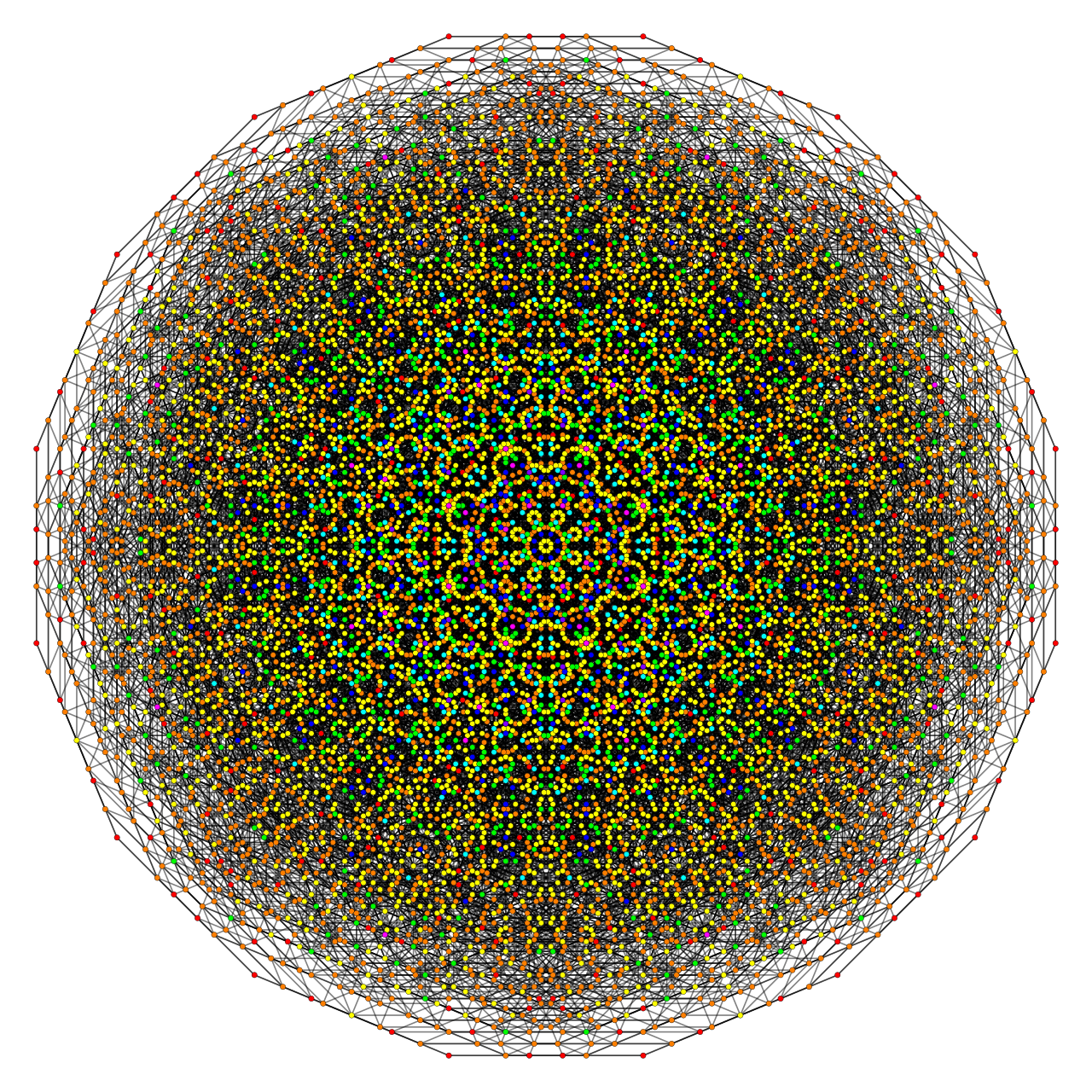}
\centering
\caption{The permutahedron of order eight, known as the omnitruncated 7-simplex, which exists in 7 dimensions and has 40,320 vertices}
\label{7simplex}
\end{figure}

There are twenty-four vertices (= 4!) associated with the permutahedron of order four (for 4D, $ {\cal N} $ = 1
supersymmetry). This same argument implies that 4D, $ {\cal N} $ = 2 theories must have a permutahedron
with 8! = 8 $\cdot$ 7  $\cdot$ 6 $\cdot$ 5  $\cdot$ 4! = (1,680)  $\cdot$ (24) = 40,320 vertices. In fact, this polytope, the Permutahedron of order 8, is known as the hexipentisteriruncicantitruncated 7-simplex, or more simply, as the omnitruncated 7-simplex, depicted in Fig. \ref{7simplex}

\be  \begin{array}{cccccccccccc}
{\bm {\cal P}}_{1} &=& \, {\bm {\rm I}}{}_{2}\ & \otimes &{\bm {\rm I}}{}_{2 
} & \otimes &{\bm {\rm I}}{}_{2}  &~=~ &()& ~=~& \langle 1 \, 2 \, 3 \, 4\, 5 \, 6 \, 7 \, 8 \,  
\rangle&~~~, ~~~ \\
{\bm {\cal P}}_{2} &=& {\bm {\rm I}}{}_{2} & \otimes &{\bm {\rm {I}}}{}_{2} & 
\otimes &{\bm \s}^1  & ~=~&  (12)(34)(56)(78)  & ~=~& 
\langle 2 \, 1 \, 4 \, 3\, 6 \, 5 \, 8 \,7 \, \rangle
&~~~, ~~~ \\
{\bm {\cal P}}_{3} &=&{\bm {\rm I}}{}_{2}  & \otimes &{\bm \s}^1 & \otimes & {\bm 
{\rm I}}{}_{2}  &~=~ &  (13)(24)(57)(68)  & ~=~& 
\langle 3 \, 4 \, 1 \, 2\, 7 \, 8 \, 5 \, 6 \, \rangle&~~~, ~~~ \\
{\bm {\cal P}}_{4} &=& {\bm {\rm I}}{}_{2} & \otimes &{\bm \s}^1 & \otimes &{\bm 
\s}^1  &~=~ &   (14)(23)(58)(67)  &~=~ &  
\langle 4 \, 3 \, 2 \, 1\, 8 \, 7 \, 6 \,5 \rangle
&~~~, ~~~ \\
{\bm {\cal P}}_{5} &=& {\bm \s}^1 & \otimes &{\bm {\rm I}}{}_{2} & \otimes &{\bm 
{\rm I}}{}_{2}  &~=~ & (15)(26)(37)(48)  & ~=~&  
\langle 5 \, 6 \, 7 \, 8\, 1 \, 2 \, 3 \, 4 \, \rangle
&~~~, ~~~ \\  
{\bm {\cal P}}_{6} &=& {\bm \s}^1 & \otimes &{\bm {\rm I}}{}_{2}  & \otimes & {\bm 
\s}^1 &~=~ &   (16)(25)(38)(47)  &~=~ & 
\langle 6 \, 5 \, 8 \, 7\, 2 \, 1 \, 4 \, 3 \, \rangle
&~~~, ~~~ \\
{\bm {\cal P}}_{7} &=& {\bm \s}^1 & \otimes &{\bm \s}^1 & \otimes &{\bm {\rm I}}{}_{2 
}  &~=~ &   (17)(28)(35)(46) & ~=~&  
\langle 7 \, 8 \, 5 \, 6\, 3 \, 4 \, 1 \, 2 \, \rangle
&~~~, ~~~ \\
{\bm {\cal P}}_{8} &=& {\bm \s}^1 & \otimes &{\bm \s}^1 & \otimes &{\bm \s}^1 
& ~=~&    (18)(27)(36)(45)  & ~=~& 
\langle 8 \, 7 \, 6 \, 5\, 4 \, 3 \, 2 \, 1 \, \rangle
&~~~. ~~~ \\
\end{array}
\label{N8diaX} \ee

Several questions remain for future inquiry. If the faces of permutahedra can be interpreted as consisting of supercharges of lower degree supermultiplets, can we embed these supercharges (of lower degree supermultiplets) of a lower degree permutahedra into higher degree permutahedra (thereby creating a mechanism for the generation of higher $\cal N$ supermultiplets from lower $\cal N$ supermultiplets? Another pressing question to pursue in the context of permutahedra is what is the interpretation of the non-closure terms in Eq.(\ref{CLFF3})? Lastly, does the Bruhat weak ordering metric on the permutations plays a role in the sorting done by the eigenvalues?

Answering these questions will require:
\newline \indent
(a.) explicit knowledge of the arrangements of {\em {all}} 40,320 permutations
       in the ${\mathbb S}{}_8$ \newline \indent $~~~~~$
       permutahedron, and \newline \indent
(b.) the calculation of all two point ``correlator'' matrices \cite{pHEDRON} between the per-
\newline \indent $~~~~~$
mutations that appear in the six representations seen above together with
 \newline \indent $~~~~~$
 those associated with the permutations described in Eq.\ (\ref{N8diaX}).
\vskip0.05in \noindent
Thanks to modern IT coding and infrastructure this is a surmountable 
problem. In a future paper these results will be reported.

\vspace{.05in}
\begin{center}
\parbox{4in}{{\it ``As the prerogative of  Natural Science is to cultivate \\ $~~$ 
a taste
for observation, that of 
Mathematics is, almost \\ $~~$ 
from the starting point, to stimulate
the faculty of \\ $~~$  invention.'' \\ ${~}$ 
${~}$ 
\\ ${~}$ }\,\,-\,
J.\  J.\ Sylvester $~~~~~~~~~$}
\parbox{4in}{
$~~$}  
\end{center}
		
\noindent
{\bf Acknowledgments}\\[.1in] \indent
This research is supported in part by the endowment of the Ford Foundation Professorship 
of Physics at Brown University and the Brown Theoretical Physics Center. Additional 
acknowledgment is given for their participation in the 2020 SSTPRS (Student Summer 
Theoretical Physics Research Session) program by Devin Bristow, John H. Caporaletti,
Aleksander Cianciara, Delina Levine, and Gabriel Yerger.

\newpage


\end{document}